\documentclass[twocolumn,showpacs]{revtex4}

\usepackage{amsmath}
\usepackage{graphicx}
\usepackage{dcolumn}
\usepackage{bm}
\usepackage{subfigure}
\usepackage{amsfonts}
\usepackage{amssymb}
\usepackage{subfigure}

\begin{document}

\preprint{ITP/UU-XXX}

\title{Inhomogeneous Superfluid Phases 	 in ${}^6$Li-${}^{40}$K mixtures at unitarity.}

\author{J.E. Baarsma}
\email{J.E.Baarsma@uu.nl}
\author{H.T.C. Stoof}

\affiliation{
Institute for Theoretical Physics, Utrecht University,\\
Leuvenlaan 4, 3584 CE Utrecht, The Netherlands}

\begin{abstract}We show that the ultracold three-dimensional ${}^6$Li-${}^{40}$K mixture at unitarity can exhibit the highly exotic Larkin-Ovchinnikov superfluid phase. We determine the phase diagram for majorities of ${}^{40}$K atoms within mean-field theory taking the inhomogeneities of the fermion states into account exactly. We find two different inhomogeneous superfluid phases in mixtures with a majority of ${}^{40}$K atoms, namely the Larkin-Ovchinnikov (LO
) phase with one inhomogeneous direction and a cubic phase (LO$^3$) where three spatial translational symmetries are broken. We determine the transition between these two phases by solving the Bogoliubov-de Gennes equations in the superfluid LO phase. Subsequently, we calculate the atomic density modulation of the atoms in the LO phase and show that it is sufficiently large to be visible in experiment.
\end{abstract}

\maketitle

Since the first realization of superfluidity in a two-component gas of fermionic atoms a large number of exciting experiments have been performed. For instance, with the use of a Feshbach resonance the inter-particle interactions can be tuned and hereby the crossover between the Bose-Einstein condensation of molecules and Cooper pairs was studied \cite{Regal,Martin}.
By inducing spin flips in a Fermi gas the number of particles in different spin states can be changed, opening up the possibility of studying the influence of a population imbalance on the phase transition to a superfluid state. This possibility has indeed been materialized in a two-component gas of ${}^6$Li atoms and the phase diagram turns out to be governed by a tricritical point below which the gas phase separates into a superfluid and normal region \cite{Ketterle,Hulet,Shin,Koos}.

One of the exotic states of matter that may also be realized in a system of ultracold Fermi atoms is a superfluid where the Cooper pairs have a nonzero center-of-mass momentum. The first to propose this possibility in the context of superfluid films in a magnetic field were Fulde and Ferrel (FF) \cite{Fulde} and independently Larkin and Ovchinnikov  (LO) \cite{Larkin}. Signatures of these so called FFLO phases have been seen in an atomic gas in one spatial dimension \cite{Liao}, but a decisive experiment that observes the FFLO correlations has not been carried out yet. In three spatial dimensions a phase with nonzero momentum Cooper pairs is predicted to be present in the population imbalanced Fermi gas with weak interactions \cite{Combescot,Yip,Leo,Bulgac,Jiang}.
However, the transition temperatures in a weakly interacting Fermi gas are very low and are expected to be out of reach with present cooling techniques.
\begin{figure}
\includegraphics[width=1.0\columnwidth]{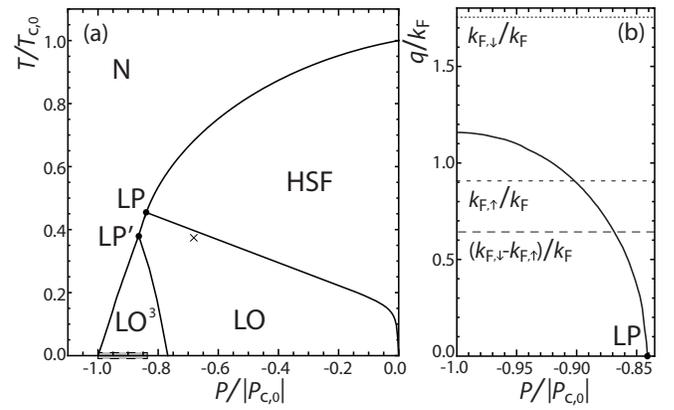}
\caption{\label{fasediagram}
Phase diagram for the unitary mixture of ${}^6$Li and ${}^{40}$K atoms (a). Each full line denotes a continuous phase transition. The grey rectangle on the horizontal axis sets the scale for panel (b), where
along the line of phase transitions to an inhomogeneous superfluid the wavevector $q$ associated with the periodicity is shown. Also shown are the Fermi wavevectors of the $^6$Li($\uparrow$) and $^{40}$K($\downarrow$) atoms and their difference, all at $T=0$ and $P=P_{c,0}$.}
\end{figure}
In the three-dimensional strongly interacting mixture of ${}^6$Li and ${}^{40}$K atoms a Lifshitz instability
towards a phase with nonzero momentum Cooper pairs is present at an attainable temperature \cite{Gubbels,Baarsma}, but the precise form of this superfluid phase has not yet been determined.
In this paper we show that at temperatures below the Lifshitz point the Fermi gas will form the exotic LO phase, which shares the properties of both a solid and a superfluid and is therefore a special kind of supersolid. This is a very exciting prospect, because even though this new kind of superfluid was first proposed by Larkin and Ovchinnikov already in 1964 \cite{Larkin}, it has never been observed in a three-dimensional fluid.
In the ${}^6$Li-${}^{40}$K mixture important steps have already been made experimentally. The Feshbach resonances of the mixture were known for some time \cite{Voigt,Walraven}, and now also expansion under the influence of strong interactions from an optical dipole trap has been realized \cite{Grimm1}, a large atom number dual-species magneto-optical trap was built \cite{Salomon}, and the repulsive polaron has been studied in this mixture \cite{Grimm2}.
At present, therefore, an ultra-cold ${}^6$Li-${}^{40}$K mixture at unitarity is to the best of our knowledge one of the most promising systems available in the laboratory to observe such a three-dimensional supersolid in a locally homogeneous trapping geometry and study its properties in detail with the accuracy of atomic physics. In addition, there recently have been proposals to observe FF and LO states in the presence of optical lattices \cite{Torma} and to observe FF states in cold atom systems with spin-orbit coupling \cite{Pu,Melo}.

The LO state is also expected to be present in a number of other condensed-matter systems, such as in superfluid $^3$He in the presence of a magnetic field \cite{Meyerovich}, neutron stars \cite{Casalbuoni} and heavy-fermion systems \cite{Matsuda}.
In the context of colorsuperconductivity a large body of theoretical work has been concerned with  different inhomogeneous phases \cite{Gatto,Mannarelli,Rajagopal,Ippolito,Mannarelli2}, all considering weak coupling and using Ginzburg-Landau expansions of the free energy. Although continuous (second-order) instabilities can be determined exactly using this expansion, the downfall is that ruling out discontinuous (first-order) transitions is impossible and one is moreover forced to remain close to the normal-superfluid transition.

In this paper we calculate for different superfluid phases where time-reversal or translational symmetry is broken the full thermodynamic potential within mean-field theory, with which both continuous and discontinuous phase transitions can be described. Moreover, in contrast to previous work, it enables us to describe transitions within the superfluid phase. In this manner we complete the phase diagram of the mixture of ${}^6$Li-${}^{40}$K atoms for a majority of ${}^{40}$K atoms, see Fig. \ref{fasediagram}(a), and especially what the structure of the inhomogeneous superfluid is. For a majority of ${}^6$Li atoms it was shown already that a tricritical point is present, below which a discontinuous phase transition towards a homogeneous superfluid, and thus phase separation, occurs \cite{Gubbels}. We find for small negative polarizations $P=(n_\uparrow-n_\downarrow)/(n_\uparrow-n_\downarrow)$, where $n_{\uparrow(\downarrow)}$ denotes the density of ${}^6$Li(${}^{40}$K) atoms, a continuous transition from the normal state (N) to a homogeneous superfluid (HSF), up to the Lifshitz point (LP). For larger majorities of ${}^{40}$K atoms there is a continuous phase transition to an inhomogeneous superfluid phase where translational symmetry is broken in one spatial direction (LO) and from the point LP' on the continuous transition is to a phase where translational symmetry is broken in three directions (LO$^3$). In Fig. \ref{fasediagram}(a) the temperature $T$ is scaled by the critical temperature $T_{c,0}$ at $P=0$ and the polarization $P$ is scaled by the critical polarization $|P_{c,0}|$ at $T=0$, because interaction effects in first instance only shift the location of the Lifshitz point \cite{Gubbels,Baarsma}, as we explain below.

This paper is organized as follows. We first explain how we calculate, within a mean-field approximation, the full thermodynamic potential for different inhomogeneous superfluid phases and how we obtain the phase diagram in Fig. \ref{fasediagram}(a). In Sec. \ref{sectie Expansion} we describe the Ginzburg-Landau expansion of the different free energies. We show that from comparing the expansion coefficients it can be directly seen which inhomogeneous superfluid phase is the most favorable. Subsequently, we discuss the possibility of having a linear position of the LO and the LO$^3$ phases and show that this does not occur.
In Sec. \ref{section Green's function} we explain how we calculate the transition line between the LO and LO$^3$ inhomogeneous phases in a direct manner. Finally, in Sec. \ref{section densities}, we determine the densities for the two atomic species in the LO phase at the conditions of the cross in Fig. \ref{fasediagram}(a), which are now position dependent with a modulation that can be as big as 10$\%$ over one period. This is therefore a convenient signature for the LO phase in experiment, but also other proposals to observe this exotic state of matter have been made \cite{Cooper,Demler}.

In this paper we calculate the phase diagram for the $^6$Li-$^{40}$K mixture within a mean-field approximation, which does not include all important physical effects such as, for instance, the screening of the interaction by particle-hole fluctuations. It is known that in the strongly interacting limit these effects play an important role and mean-field theory does not give good quantitative results. However, the phase diagram for the unitary Fermi mixture with population imbalance predicted by mean-field theory was reproduced qualitatively by renormalization-group calculations that do incorporate fluctuation and interaction effects \cite{Koos} and, more importantly, that are in agreement with the phase diagram mapped out experimentally \cite{Ketterle,Hulet,Shin}. From this it can be concluded that for these purposes the mean-field calculation already contains the relevant physics, even at unitarity. The Lifshitz point found in the $^6$Li-$^{40}$K mixture using mean-field theory remains present when adding screening and selfenergy effects and its position only changes quantitatively \cite{Gubbels,Baarsma}. This qualitative succes of mean-field theory is, physically, due to the
fact that even at unitarity the atomic selfenergies are well approximated
by a momentum and frequency independent constant. As a result,
thermodynamic instabilities are determined by mean-field-like correlation
functions with strongly renormalized constants that lead to quantitative
shifts only in the transition lines. In conclusion, mean-field theory is even at unitarity a good first approximation to explore the superfluid phases that can occur in the $^6$Li-$^{40}$K mixture and map out the phase diagram, which is the main aim of this paper.

\section{Thermodynamic potential}
Phase transitions can be determined by studying the Landau free energy of a system as a function of the appropriate order parameter. Here, the role of the Landau free energy is played by the grand-canonical thermodynamic potential $\Omega(\Delta)$, where $\Delta$, the gap parameter describing the condensate of Cooper pairs, is the order parameter for the phase transition from the  normal state to a superfluid.

We can calculate the thermodynamic potential $\Omega$ for the two-component Fermi gas with mass and population imbalance from the microscopic action
\begin{align}
\nonumber S=\int d{\bf x}d\tau\Bigg\{\sum_{\sigma=\uparrow,\downarrow} \phi_\sigma^*\left(\hbar\frac{\partial}{\partial\tau}-\frac{\hbar^2\nabla^2}{2m_\sigma}-\mu_\sigma\right)\phi_\sigma\\
-\frac{|\Delta|^2}{V_0}+\phi^*_\uparrow\phi^*_\downarrow\Delta+\Delta^*\phi_\downarrow\phi_\uparrow\Bigg\},
\label{action}
\end{align}
where $\phi_{\sigma}({\bf x},\tau)$ are Grassmann-valued fields describing fermions. In the case under study, $\uparrow$($\downarrow$) denotes a $^6$Li($^{40}$K) atom with mass $m_{\uparrow(\downarrow)}$. A population imbalance between the two atomic species is introduced by having different chemical potentials $\mu_\sigma$ for the two species, resulting in different densities, since $n_\sigma=-\partial\Omega/\partial\mu_\sigma$. The microscopic interaction strength between the fermionic atoms is denoted by $V_0$ and is attractive here. The last two terms in the above action represent respectively the annihilation and creation of a Cooper pair, described by the bosonic complex pairing field $\Delta({\bf x},\tau)$, consisting here of a ${}^6$Li and a ${}^{40}$K atom.

From the action the partition function can be
calculated as $Z=\int\text{d}[\phi^*]\text{d}[\phi]\text{d}[\Delta^*]\text{d}[\Delta]\exp(-S[\phi,\Delta]/\hbar)$ and subsequently the thermodynamic potential can be
obtained via $\Omega=-\log Z/\beta$, where $1/\beta = k_BT$ is the
inverse thermal energy with $k_B$ Boltzmann's constant.

In the partition function $Z$ not all path integrals can be evaluated exactly and therefore we use a mean-field approximation to calculate the thermodynamic potential. In this approximation the pairing field is replaced by
its most probable value and the associated path integral omitted. In practice the thermodynamic potential is then obtained by making an ansatz for the pairing field and thus for the superfluid phase. The thermodynamic potential describing the phase transition
from a normal gas to a homogeneous BCS superfluid, with both time-reversal and translational symmetry present, is calculated by taking for the pairing field $\Delta({\bf x},\tau) =\Delta_0$. Thus, a Cooper pair consists here of two fermions with opposite momentum, such that the pair has no net momentum.

Fulde and Ferrel considered a plane wave for the pairing field, $\Delta_\text{FF}({\bf x},\tau) = \Delta_0e^{i{\bf q\cdot x}}$, in which case a Cooper pair has net momentum ${\bf q}$. The superfluid described hereby still obeys translational symmetry but time-reversal symmetry is broken.
After replacing the pairing field in Eq. (\ref{action}) by the FF ansatz and using a Matsubare expansion for the fermionic fields, 
\begin{align}
\phi_\sigma({\bf x},\tau)=\sum_{{\bf k},n}\phi_{\sigma,{\bf k},n}\frac{e^{i({\bf k\cdot x}-i\omega_n\tau)}}{\sqrt{\hbar\beta V}},
\end{align}
where the summation over $n$ runs over the odd Matsubara frequencies $\omega_n$, the terms describing the creation and annihilation of a Cooper pair read
\begin{align}
\sum_n\sum_{\bf k}\bigg\{\phi_{\uparrow,{\bf k},n}\phi_{\downarrow,{\bf -k+q},-n}\Delta_0+\text{c.c.}\bigg\}.
\end{align}
The action can now be rewritten using matrix multiplication after which it reads
\begin{align}
\nonumber S=-\hbar\beta V\frac{|\Delta_0|^2}{V_0}+\sum_{\bf k}\Bigg\{\hbar\beta(\varepsilon_{\downarrow,{\bf -k+q}}-\mu_\downarrow)\\
+\sum_n\Phi^\dagger\left(\begin{array}{cc}-i\hbar\omega_n+\xi_{\uparrow,{\bf k}}&\Delta_0\\\Delta_0&-i\hbar\omega_n-\xi_{\downarrow,{\bf -k+q}}\end{array}\right)\Phi_\Bigg\},
\label{actiematrix}
\end{align}
where  $\Phi_{\bf k,n}^\dagger=[\phi_{\uparrow,{\bf k},n},\phi^*_{\downarrow,{\bf -k+q},-n}]$ and $V$ is the volume. The kinetic energy of a fermionic atom in state $|\sigma\rangle$ is $\varepsilon_{\sigma,{\bf k}}=\hbar^2 k^2/2m_\sigma$ and $\xi_{\sigma,{\bf k}}=\varepsilon_{\sigma,{\bf k}}-\mu_\sigma$. The above action can be diagonalized analytically by making a Bogoliubov transformation. Subsequently, the integral over the fermionic fields can be evaluated, after which the thermodynamic potential reads
\begin{align}
\nonumber\frac{\Omega_{\text{FF}}(\Delta_0,q)}{V}&=\frac1V\int\frac{\text d \bf k}{(2\pi)^3}\Bigg\{\frac{|\Delta_0|^2}{2\varepsilon_{\bf k}}+\omega_{\bf k,q}-\sqrt{\omega_{\bf k,q}^2+|\Delta_0|^2}\\
&-\frac{1}{\beta}\sum_{\sigma}\log\left[1+e^{-\beta\hbar\omega_{\sigma,{\bf k,q}}}\right]\Bigg\}-\frac{|\Delta_0|^2}{T^{2{\text B}}(0)},
\label{omegaFF}
\end{align}
where we replaced the summation over ${\bf k}$ by an integral, because we are interested in the thermodynamic limit. The kinetic energy $\varepsilon_{\bf k}=\hbar^2k^2/2m$ is associated with twice the reduced mass, $m=2m_\uparrow m_\downarrow/(m_\uparrow+m_\downarrow)$, and the dispersions describing the Bogoliubov quasiparticles read
\begin{figure}
\includegraphics[width=\columnwidth]{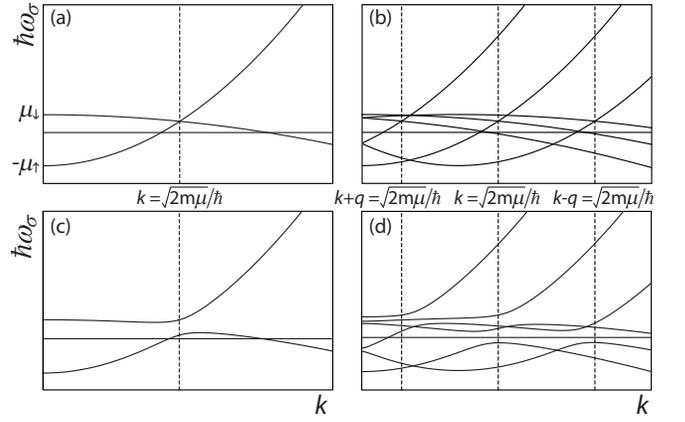}
\caption{Dispersions for the Bogoliubov quasiparticles in the homogeneous case, panel (a) and (c), and in the LO case, panel (b) and (d), for ${\bf k}$ parallel to the lattice wavevector ${\bf q}$. The upper panels are dispersions of the Fermi gas in the normal state, where the gap is zero $\Delta_0=0$. The lower two panels depict dispersions with a nonzero gap $\Delta_0$}
\label{dispersies}
\end{figure}
\begin{align}
\hbar\omega_{\sigma,{\bf k,q}}&=\sqrt{\omega_{\bf k,q}^2+|\Delta_0|^2}+\frac{\sigma}{2}[\varepsilon_{\uparrow,{\bf k}}-\varepsilon_{\downarrow,{\bf -k+q}}-2h],
\label{dispersions}
\end{align}
with $\omega_{\bf k,q}=[\varepsilon_{\uparrow,{\bf k}}+\varepsilon_{\downarrow,{\bf -k+q}}-2\mu]/2$, where $\mu=(\mu_\uparrow+\mu_\downarrow)/2$ is the average chemical potential. Above, $\sigma=+1(-1)$ for $\uparrow(\downarrow)$ atoms and $h=(\mu_\uparrow-\mu_\downarrow)/2$ is the chemical potential difference.
In the thermodynamic potential we replaced the microscopic interaction $V_0$ by the two-body transition matrix $T^{2{\text B}}$
\begin{align}
\frac{1}{V_0}=\frac{1}{T^{2{\text B}}(0)}-\frac1V\sum_{\bf k}\frac1{2\varepsilon_{\bf k}},
\end{align}
which is related to the $s$-wave scattering length $a$, $1/T^{2{\text B}}(0)=m/4\pi\hbar^2a$. Experimentally, the scattering length can be controlled by the use of a Feshbach resonance. In this paper we consider the unitarity regime, where $1/|a|=0$.

The above thermodynamic potential describes the transition from a normal gas to the homogeneous BCS superfluid, with time-reversal symmetry, when the momentum ${\bf q}$ of the Cooper pairs is set to zero, i.e.,  $\Omega_\text{BCS}(\Delta_0)=\Omega_\text{FF}(\Delta_0,0)$. Correspondingly, for ${\bf q}=0$ the dispersions in Eq. (\ref{dispersions}) describe the Bogoliubov quasiparticles in the homogeneous superfluid, see Fig. \ref{dispersies}.

Larkin and Ovchinnikov made an ansatz for the Cooper pair that results in a truly inhomogeneous superfluid, but with time-reversal symmetry unbroken. Namely, they assumed a standing wave,
\begin{align}
\Delta_{\text{LO}}({\bf x},\tau) = \frac{\Delta_0}{\sqrt2}\left[e^{i{\bf q\cdot x}}+e^{-i{\bf q\cdot x}}\right],
\label{LO1ansatz}
\end{align}
in which case not only the Cooper-pair phase but also the superfluid density becomes position dependent. Plugging the LO ansatz in Eq. (\ref{action}) results in the following expression for the Cooper-pair terms
\begin{align}
\nonumber\frac1{\sqrt2}\sum_n&\sum_{\bf k}\bigg\{\phi_{\uparrow,{\bf k},n}\phi_{\downarrow,{\bf -k+q},-n}\Delta_0+\text{c.c.}\\
&+\phi_{\uparrow,{\bf k},n}\phi_{\downarrow,{\bf -k-q},-n}\Delta_0+\text{c.c.}\bigg\},
\label{LOcouplings}
\end{align}
where an important difference with the previous ansatz is that via the pairing field now every fermionic field component $\phi_{\sigma,{\bf k},n}$ couples to two other components, $\phi_{-\sigma,{\bf -k+q},-n}$ and $\phi_{-\sigma,{\bf -k-q},-n}$, instead of only to $\phi_{-\sigma,{\bf -k+q},-n}$. This has important consequences for the computation of the thermodynamic potential $\Omega_{\text{LO}}$. Namely, rewriting the action using matrix multiplication here not yields a $2\times2$ matrix but one with infinite dimensions. In order to perform a calculation we truncate this matrix and thus neglect some couplings. The smallest matrix we consider has dimension $D=6$ and after diagonalizing we find six dispersions $\hbar\omega_i$, see Fig. \ref{dispersies}. Notice that if the gap is zero, Fig. \ref{dispersies}(a) and (b), the dispersions are the same as the homogeneous dispersions, only displaced by the lattice wavevector ${\bf q}$. While, if there is a nonzero gap, Fig. \ref{dispersies}(c) and (d), the LO case is truly different from the homogeneous superfluid since a band structure appears.
\begin{figure}
\includegraphics[width=\columnwidth]{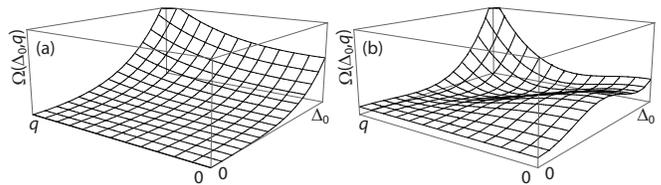}
\caption{The thermodynamic potential for the LO configuration, as a function of the amplitude of the gap $\Delta_0$ and the lattice wavevector $q$. In panel (a) the Fermi gas is in the normal state, whereas in panel (b) a transition to an inhomogeneous superfluid has occured. Namely, the global minimum of the thermodynamic potential in panel (b) lies at nonzero $\Delta_0$ and $q$.}
\label{OmegaLO}
\end{figure}

With the LO quasiparticle dispersions the thermodynamic potential $\Omega_{\text{LO}}$ can be calculated, where attention should be paid to a few differences with the homogeneous calculation. Firstly, if a $D\times D$ dimensional matrix is used then $\Omega_{\text{LO}}$ contains a summation over $D$ dispersions, while we only describe two species of fermions. To compensate for this, we have to multiply the summation by $2/D$.

Secondly, the integrand in the thermodynamic potential needs to be convergent in order to be able to evaluate the integral over momentum. In the homogeneous case, and also in the FF case, there are several  divergencies in the integrand that exactly cancel against each other in the limit $k\rightarrow\infty$. In particular, the divergency $-m|\Delta_0|^2/k^2$, coming from the square root in Eq. (\ref{omegaFF}) and originating from diagonalizing the matrix in Eq. (\ref{actiematrix}), cancels against $|\Delta_0|^2/2\varepsilon_{\bf k}$ originating from replacing the microscopic interaction $V_0$ by the $s$-wave scattering length $a$.

In the present LO case, the latter divergency enters the thermodynamic potential in the same manner, while the matrix to be diagonalized differs and moreover is truncated, whereby some couplings are neglected. This truncation causes the two divergencies to not cancel in the limit $k\rightarrow\infty$ and we need to account for this. If the matrix has dimension $D$, there are $D$ momenta involved and there should thus be $2D$ couplings, see Eq. (\ref{LOcouplings}). However, there are only $2D-4$ coupling terms accounted for after the truncation, which means that every fermionic field component couples on average to $(2D-4)/D$ other fermionic components. We multiply the term originating from replacing $V_0$ by $a$ by this number divided by two, such that the divergencies cancel again and the integral can be evaluated. Note that in the limit $D\rightarrow\infty$ this prefactor is equal to 1 again, $(2D-4)/2D\rightarrow1$.

Finally, the thermodynamic potential should not depend on the dimension of the matrix. We calculated the thermodynamic potential for increasing matrix size to check convergence. At the phase transition convergence is rapid, because we find a second-order phase transition and the gap $\Delta_0$ is small, while we need to go to larger matrix dimensions inside the superfluid phase.

The thermodynamic potential for the LO case is plotted in Fig. \ref{OmegaLO} for different chemical potentials and temperatures. In panel (a) the Fermi gas is in the normal state, whereas in panel (b) a phase transition has occurred. There the global minimum is located at nonzero gap and lattice wavevector $q$ and it can be seen that the global minimum shifted from zero in a continuous fashion to this position.

\begin{figure}
\includegraphics[width=1.0\columnwidth]{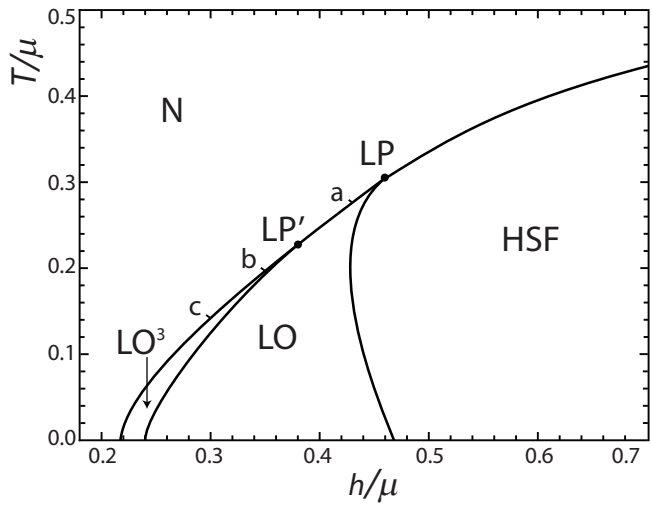}
\caption{\label{fasediagram2}
Phase diagram for the unitary mixture of ${}^6$Li and ${}^{40}$K atoms as a function of temperature $T$ and the chemical potential difference $h$, both scaled by the average chemical potential $\mu$. Each full line denotes a continuous phase transition.}
\end{figure}

Obviously, many other choices for the Cooper-pair wavefunction can be made, which all result, by following the steps described above, in a different thermodynamic potential describing a particular phase transition.
We calculated the thermodynamic potential for different ansatzes and find for real wavefunctions, such as the LO ansatz, higher critical temperatures than for complex wavefunctions, such as the FF ansatz, which means that phases where time-reversal symmetry is unbroken are more favorable.
We compared different configurations of standing waves for the Cooper-pair wavefunction. Namely, we calculated $\Omega$ for two (LO$^2$) and three (LO$^3$) perpendicular standing waves with equal amplitudes, where the Cooper-pair ansatzes are
\begin{align}
\Delta_{\text{LO}^2}({\bf x},\tau) &= \frac{\Delta_0}{\sqrt4}\left[e^{iqz}+e^{-iqz}+e^{iqx}+e^{-iqx}\right],\\
\nonumber\Delta_{\text{LO}^3}({\bf x},\tau) &= \frac{\Delta_0}{\sqrt6}\left[e^{iqz}+e^{-iqz}\right.\\
&\left.+e^{iqx}+e^{-iqx}+e^{iqy}+e^{-iqy}\right],
\end{align}
and for a triangular and tetrahedral configuration of standing waves with equal amplitudes. By comparing the thermodynamic potentials we completed the phase diagram for the Fermi mixture of ${}^{6}$Li and  ${}^{40}$K atoms, see Figs. \ref{fasediagram}(a) and \ref{fasediagram2}. We find that the transition from the normal gas to the inhomogeneous superfluid phase is continuous, i.e., of second order, independently of the structure of the inhomogeneous phase. For all configurations we find the same critical temperature.
The lattice wavevector, i.e., the value of $q$ for which the thermodynamic potential $\Omega(\Delta_0,q)$ has a global minimum, along the line of the phase transition is shown in Fig. \ref{fasediagram}(b). It can be seen that for $T=0$ it differs from the difference in Fermi wavevectors, which is due to the fact that we are in the strongly interacting regime. We found that inside the LO superfluid phase the global minimum of the thermodynamic potential shifts continuously from a nonzero to a zero lattice wavevector $q$, in which case the superfluid is homogeneous again. This criterion for the transition between the inhomogeneous and homogeneous superfluid is equivalent to the vanishing of the domain-wall energy \cite{Yip,Leo}.

\section{Ginzburg-Landau Expansion}
\label{sectie Expansion}
Since we have just found that the phase transition from the normal gas to the inhomogeneous superfluid is continuous we can use a Ginzburg-Landau expansion of the thermodynamic potential
\begin{align}
\Omega(\Delta_0,q)\simeq\alpha(q)|\Delta_0|^2+\frac{\beta(q)}2|\Delta_0|^4+\ldots.
\label{Landauexpansion}
\end{align}
A second-order phase transition occurs the moment $\alpha(q)$ changes sign. The coefficients in the above expansion can be calculated by taking derivatives of the thermodynamic potential, for example
\begin{align}
\alpha(q)=\left.\frac{\partial\Omega(\Delta_0,q)}{\partial|\Delta_0|^2}\right|_{\Delta_0=0}.
\end{align}
Equivalently, the coeffecients can be determined by calculating the amplitude of the proper Feynman diagram \cite{Kleinert}. The advantage of using Feynman diagrams instead of derivatives here is that the former is an exact linear-response calculation in the normal state where we do not have to solve a matrix problem of infinite dimensions.

\begin{figure}
\begin{center}
\includegraphics[width=.4\columnwidth]{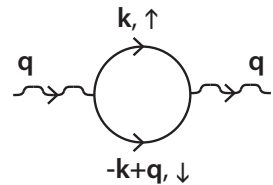}
\end{center}
\caption{Feynman diagram. Wiggly lines denote pairing fields and straight lines denote the fermionic propagators.}
\label{alpha}
\end{figure}

The second-order coefficient $\alpha$ is related to the Feynman diagram shown in Fig. \ref{alpha}. The external legs correspond to pairing fields $\Delta$, while the internal lines represent fermionic propagators $\phi_\sigma$. The amplitude of this diagram is given by
\begin{align}
\frac{1}{\hbar\beta}\sum_n\int\frac{\text{d}^3{\bf k}}{(2\pi)^3} G_{0;\uparrow}({\bf k},\text{i}\omega_n)G_{0;\downarrow}({\bf -k+q},-\text{i}\omega_n),
\end{align}
where $G_{0;\sigma}^{-1}({\bf k},\text{i}\omega_n)=-{\text i}\omega_n+\varepsilon_{\sigma,{\bf k}}-\mu_\sigma$ is the inverse non-interacting Green's function describing a fermionic atom in state $|\sigma\rangle$. After splitting the fractions and summing over the Matsubara frequencies it reads
\begin{align}
\nonumber\alpha(q)=&\int\frac{\text{d}{\bf k}}{(2\pi)^3}\left\{\frac{1}{2\varepsilon_{\bf k}}+\right.\\
&\left.\frac{N_{\text F}(\xi_{\uparrow,{\bf k}})+N_{\text F}(\xi_{\downarrow,{\bf -k+q}})-1}{\xi_{\uparrow,{\bf k}}+\xi_{\downarrow,{\bf -k+q}}}\right\},
\end{align}
where the first term originates from the term quadratic in $\Delta$ in Eq. (\ref{action}) and the second term is the amplitude of the Feynman diagram. The Fermi distribution functions are given by $N_{\text F}(x)=1/[\exp(\beta x)+1]$.
The quadratic coefficient for a given lattice configuration is now obtained by summing the above expression over the lattice wavevectors and multiplying it by the normalization squared of $\Delta({\bf x})$, which in the LO case yields for example
\begin{align}
\alpha_{\text{LO}}(q)=\frac{1}{2}[\alpha(q)+\alpha(-q)]=\alpha(q),
\end{align}
because this ansatz contains two momenta, ${\bf q}$ and ${\bf -q}$. Determining this coefficient for other wavefunctions, leads to the conclustion that the $\alpha$ coefficient is the same for every lattice configuration, which confirms our findings from the full thermodynamic potentials. Namely, we find a continuous phase transition at a critical temperature independent of the lattice configuration.
Thus, to compare the different lattice configurations we need to look at the fourth-order coefficient $\beta(q)$ in the expansion in Eq. (\ref{Landauexpansion}).

\begin{figure}
\begin{center}
\includegraphics[width=\columnwidth]{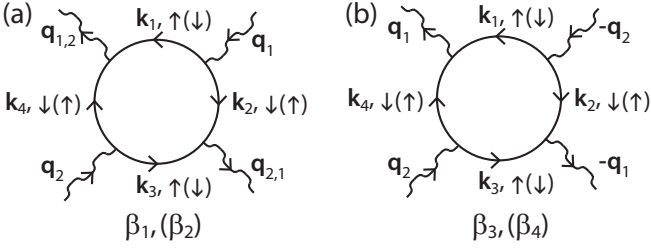}
\end{center}
\caption{The fourth-order diagrams of which we need to calculate the amplitude. External momenta are nonzero, while the external bosonic Matsubara frequencies are equal to zero. The Matsubara frequencies are not shown in this figure.}
\label{betadiagram1}
\end{figure}

This coefficient can be calculated from the amplitudes of all the diagrams with four external legs, shown in Fig. \ref{betadiagram1}. In the case of nonzero external momenta and in the presence of a mass and population imbalance there are four distinct diagrams with four external legs, which are denoted by $\beta_i$ in Fig. \ref{betadiagram1}. Without external momentum all these diagrams give the same amplitude. Without mass and population imbalance there are only two distinct diagrams, namely in the balanced case $\beta_1=\beta_2$ and $\beta_3=\beta_4$. Moreover, $\beta_{3,4}$ are only different from $\beta_{1,2}$ if the angle between the two external momenta is not equal to zero or $\pi$.
In all lattice configurations we consider the lengths of the lattice wavevectors are the same, which means that $|{\bf q_1}|=|{\bf q_2}|=q$, because this minimizes the quadratic part $\alpha(q)$ of the Ginzburg-Landau expansion. The amplitudes then only depend on the magnitude $q$ and the angle between the external momenta, i.e., $\beta({\bf q_1,q_2})=\beta(q,\theta)$.
The amplitude of the diagrams in Fig. \ref{betadiagram1}(a) are given by
\begin{align}
\nonumber&\beta_{1,2}(q,\theta)=\frac{1}{\hbar\beta}\sum_n\int\frac{\text{d}{\bf k}}{(2\pi)^3}
G_{0;\sigma}({\bf k}+{\bf q}_1-{\bf q}_2,\text{i}\omega_n)\\
&\cdot G_{0;-\sigma}({\bf q}_2-{\bf k},-\text{i}\omega_n)G_{0;\sigma}({\bf k},\text{i}\omega_n)G_{0;-\sigma}({\bf q}_2-{\bf k},-\text{i}\omega_n),
\end{align}
where we already used conservation of frequency and momentum at all vertices. Again fractions can be split and the sum over the Matsubara frequencies performed. The resulting expression becomes rather lengthy and is omitted here. The diagrams in Fig. \ref{betadiagram1}(b) yield similar expressions, namely
\begin{align}
\nonumber&\beta_{3,4}(q,\theta)=\frac{1}{\hbar\beta}\sum_n\int\frac{\text{d}{\bf k}}{(2\pi)^3}
G_{0;\sigma}({\bf k}+{\bf q}_1-{\bf q}_2,\text{i}\omega_n)\\
&\cdot G_{0;-\sigma}(-{\bf q}_1-{\bf k},-\text{i}\omega_n)G_{0;\sigma}({\bf k},\text{i}\omega_n)G_{0;-\sigma}({\bf q}_2-{\bf k},-\text{i}\omega_n).
\end{align}
To obtain the fourth-order coefficient for a given lattice configuration we sum the relevant diagrams and multiply with the normalization factor to the power 4. The expression for $\beta(q)$ for the LO ansatz reads
\begin{align}
\beta_{\text{LO}}(q)=\frac14\Big[2\beta_{1+2}(q,0)+4\beta_{1+2}(q,\pi)\Big],
\label{betalo}
\end{align}
where $\beta_{1+2}$ is short-hand notation for $\beta_1+\beta_2$. Here, if the external momenta are equal in $\beta_{1,2}$, ${\bf q_1=q_2}$, there are two possibilities for ${\bf q_1}$, namely ${\bf q}$ and ${\bf -q}$, which explains the factor 2 in front of $\beta_{1+2}(q,0)$ above. In the case of unequal external momenta there are four possibilities for $\beta_{1,2}$. Namely, there are two choices for ${\bf q_1}$, after which ${\bf q_2}$ is fixed. Subsequently, there are still two possibilities for the outgoing momenta, denoted by ${\bf q_{1,2}}$ in Fig. \ref{betadiagram1}(a), which makes 4 possibilities in total. In the LO coefficient $\beta_{3,4}$ does not occur.
In the case of the LO$^2$ ansatz the expression for $\beta(q)$ reads
\begin{align}
\nonumber\beta_{\text{LO}^2}(q)=&\frac1{16}\Big[4\beta_{1+2}(q,0)+8\beta_{1+2}(q,\pi)\\
&+16\beta_{1+2}(q,\frac\pi2)+8\beta_{3+4}(q,\frac\pi2)\Big],
\end{align}
where the numerical factors are obtained using the same reasoning as above. The LO$^3$ coefficient is given by
\begin{align}
\nonumber\beta_{\text{LO}^3}(q)=&\frac1{36}\Big[6\beta_{1+2}(q,0)+12\beta_{1+2}(q,\pi)\\
&+48\beta_{1+2}(q,\frac\pi2)+24\beta_{3+4}(q,\frac\pi2)\Big].
\end{align}
Finally, the triangular configuration leads to the expression
\begin{align}
\nonumber\beta_{\text{triangular}}(q)=&\frac1{36}\Big[6\beta_{1+2}(q,0)+12\beta_{1+2}(q,\pi)\\
\nonumber &+24\beta_{1+2}(q,\frac\pi3)+24\beta_{1+2}(q,\frac{2\pi}3)\\
&+24\beta_{3+4}(q,\frac\pi3)\Big],
\end{align}
while the tetrahedral configuration yields
\begin{align}
\nonumber\beta_{\text{tetrahedral}}(q)=&\frac1{48}\Big[8\beta_{1+2}(q,0)+16\beta_{1+2}(q,\pi)\\
\nonumber &+48\beta_{1+2}(q,\phi_0)+48\beta_{1+2}(q,\pi-\phi_0)\\
&+48\beta_{3+4}(q,\phi_0)\Big],
\end{align}
where $\phi_0=\arccos(-1/3)$.

\begin{figure}
\begin{center}
\includegraphics[width=\columnwidth]{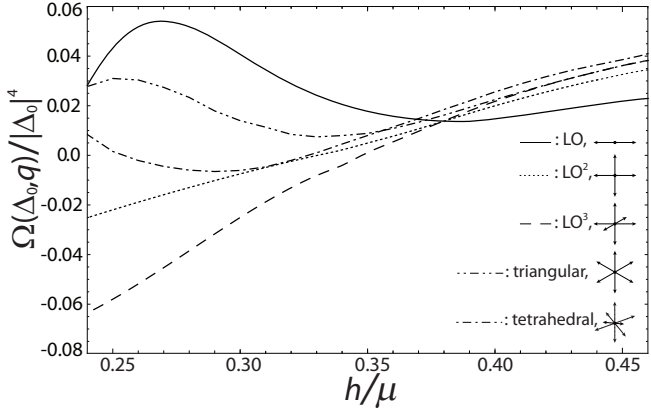}
\end{center}
\caption{The thermodynamic potential along the normal to superfluid transition line from Fig. \ref{fasediagram2}(a). At first the LO (solid line) gives the lowest thermodynamic potential, while for larger population imbalances this is true for the LO$^3$ phase (dashed line). An LO$^2$ phase (dotted) or more complicated configurations (dashed-dotted and dashed-double-dotted) never form the equilibrium state of this system.}
\label{Omegags}
\end{figure}

We calculate the thermodynamic potential in Eq. (\ref{Landauexpansion}) along the line of second-order phase transitions from the normal gas to the inhomogeneous superfluid phase. The result is shown in Fig. \ref{Omegags} and it can be seen that at the Lifshitz point the LO phase gives the lowest energy and is thus the most favorable lattice configuration, whereas for larger majorities of $^{40}$K atoms it becomes more favorable to break translational symmetry in three spatial directions. Namely, then the LO$^3$ configuration gives the lowest thermodynamic potential.

\begin{figure}
\begin{center}
\includegraphics[width=\columnwidth]{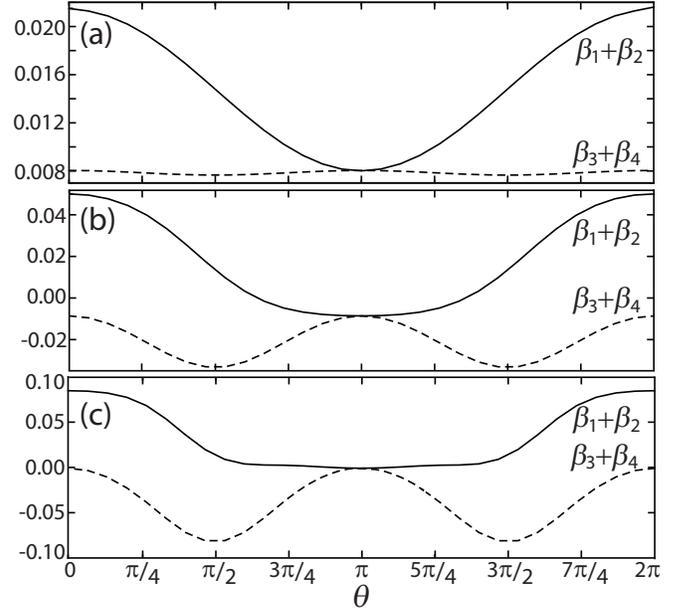}
\end{center}
\caption{Amplitudes of the fourth-order diagrams as a function of the angle $\theta$ between ${\bf q_1}$ and ${\bf q_2}$. In all panels the full line is the sum of $\beta_{1}$ and $\beta_2$ and the dashed line is $\beta_3+\beta_4$. Panel (a), (b) and (c) depict the amplitudes at the conditions marked by a, b and c in Fig. \ref{fasediagram2} respectively.}
\label{hoeken}
\end{figure}

In order to better understand this result, we now look at the dependence of the amplitudes of the Feynman diagrams in Fig. \ref{betadiagram1} on the angle $\theta$ between ${\bf q_1}$ and ${\bf q_2}$. The angular dependencies are shown in Fig. \ref{hoeken}, for three points along the line of phase transitions.
The amplitudes in Fig. \ref{hoeken}(a) are calculated at the conditions marked by a in Fig. \ref{fasediagram2}, close to the Lifshitz point. Panel (b) corresponds to lower temperatures, just below the transition to the LO$^3$ phase. It can be seen that energy is now gained by having $\pi/2$ angles in the Cooper-pair wavefunction. This explains why it is now favorable to break translational symmetry in all directions, instead of in only one direction as in the LO phase. For lower temperatures, as in panel (c), this energy gain is even larger.
These plots also clarify why more complicated Cooper-pair wavefunctions, such as the triangular configuration, never form the ground state of the system. Namely, these configurations include different angles than the angles for which the amplitudes of the fourth-order diagrams are minimal.

\subsection{Continuous transition from LO to LO$^3$}
\label{continuous transition}
\begin{figure}
\includegraphics[width=\columnwidth]{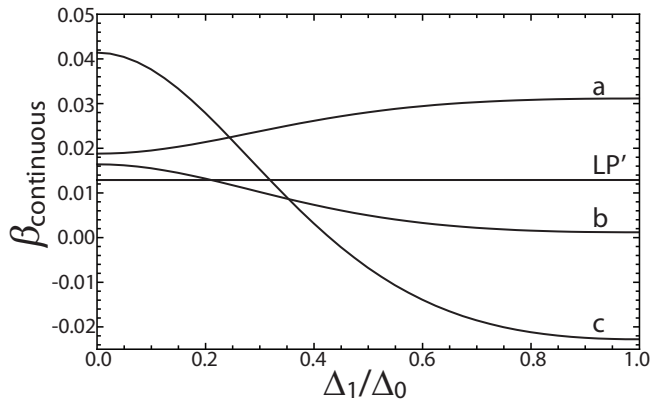}
\caption{The continuous ansatz as a function of the perpendicular amplitude $\Delta_1$ divided by the LO amplitude $\Delta_0$ for different points along the line of continuous phase transitions. The letters a, b and c correspond to the conditions marked in Fig. \ref{fasediagram2}. The horizontal line corresponds to the point LP' in Fig. \ref{fasediagram2} where the LO and LO$^3$ phase meet.}
\label{betavana}
\end{figure}

A possibility we did not consider so far is a transition from the LO to the LO$^3$ phase, where the two standing waves perpendicular to the LO standing wave have an amplitude that continuously changes from zero for higher temperatures to a nonzero value for temperatures below LP'. The Cooper-pair ansatz describing this possibility has the following form
\begin{align}
\nonumber\Delta({\bf x})=&\frac1{\sqrt{2+4p^2}}\Big[\Delta_0(e^{i q_x x}+e^{-i q_x x})\\
&+\Delta_1(e^{i q_y y}+e^{-i q_y y}+e^{i q_z z}+e^{-i q_z z})\Big],
\end{align}
where $p=\Delta_1/\Delta_0$ and the perpendicular amplitude $\Delta_1$ can now change continuously from zero to some nonzero value. Also for this ansatz the quadratic coefficient $\alpha(q)$ in Eq. (\ref{Landauexpansion}) has the same form and the fourth-order coefficient reads
\begin{align}
\nonumber\beta_{\text{continuous}}(q)=&\frac1{(2+4p^2)^2}\Big[(2+4a^2)\beta_{1+2}(q,0)\\
\nonumber&+(4+8p^2)\beta_{1+2}(q,\pi)\\
\nonumber&+(32p^2+16p^4)\beta_{1+2}(q,\frac\pi2)\\
&+(16p^2+8p^4)\beta_{3+4}(q,\frac\pi2)\Big],
\end{align}
which reduces to $\beta_{\text{LO}}$ for $p=0$, when $\Delta_1=0$, and to $\beta_{\text{LO}^3}$ for $p=1$, when $\Delta_1=\Delta_0$.
In Fig. \ref{betavana} we plot the above coefficient as a function of the perpendicular amplitude for different points along the line of continuous phase transitions from the normal to the inhomogeneous superfluid state. The minimum of this coefficient determines the inhomogeneous superfluid that is the equilibrium state. For point a in Fig. \ref{fasediagram2} the minimum is located at $\Delta_1=0$, see Fig. \ref{betavana}, which means that the equilibrium state of the Fermi mixture for these conditions is the LO state. While, for points b and c the LO$^3$ phase occurs and at the transition point LP' the energy of the two phases is equal. We conclude that between the regions where respectively the LO and LO$^3$ phase occur there is not a region where translational symmetry is already broken in three spatial directions but with a smaller amplitude in the directions perpendicular to the LO direction.

\section{LO Green's Function}
\label{section Green's function}
We found that for one point (LP') on the normal to superfluid transition line the thermodynamic potentials for the LO and the LO$^{3}$ phase are equal, see Figs. \ref{fasediagram2} and \ref{Omegags}. On the high-temperature side of LP' the transition from the normal gas is to an inhomogeneous superfluid with one broken spatial symmetry, while for lower temperatures it is to a phase with three broken spatial symmetries. Within the superfluid region the transition line between these two different inhomogeneous phases can be obtained by again comparing thermodynamic potentials, but now that we know which phases to compare we can also calculate the transition line more directly. In order to do so, we split the order parameter for the LO$^3$ phase as
\begin{align}
\Delta_{\text{LO}^3}({\bf x})=\Delta_{\text{LO}}(z)+\Delta_{\perp}[\cos(qx)+\cos(qy)],
\end{align}
where $\Delta_{\text{LO}}(z)$ is given in Eq. (\ref{LO1ansatz}). We now expand the thermodynamic potential for the cubic phase in the perpendicular order parameter $\Delta_\perp$
\begin{align}
\Omega_{\text{LO}^3}\simeq\Omega_{\text{LO}}(\Delta_0,q)+\alpha_\perp(\Delta_0,q)|\Delta_\perp|^2.
\end{align}
At the moment $\alpha_\perp$ changes sign the minimum of $\Omega_{\text{LO}^3}$ shifts to a nonzero value of $\Delta_\perp$, which means that the LO$^3$ phase is more favorable than the LO phase.
In order to calculate $\alpha_\perp$ we need to determine the Green's function for the LO phase, since
\begin{align}
\alpha_\perp=-\frac{V}{V_0}+\frac{k_BT}{2|\Delta_\perp|^2}\text{Tr}\left[{\bf G}_{\text{LO}}{\bf \Delta}_\perp{\bf G}_{\text{LO}}{\bf \Delta}_\perp\right],
\end{align}
where ${\bf \Delta}_\perp(x,y)=(\Delta_\perp[\cos(qx)+\cos(qy)]/\hbar)\sigma_x$, with $\sigma_x$ the first Pauli matrix in Nambu space. The trace in the above equation is taken over real space, imaginary time and Nambu space. The inverse Green's function ${\bf G}_{\text{LO}}^{-1}$ is known from Eq. (\ref{action}) and can be expanded in its energy eigenmodes, which are the Bogoliubov quasiparticle wavefunctions. In this manner we obtain an expression for the Green's function in the LO phase, which contains the inhomogeneities of this phase in an exact way. The quasiparticle wavefunctions for the Larkin-Ovchinnikov phase we calculate by solving the Bogoliubov-de Gennes equation
\begin{align}
\left(\begin{array}{cc}-\frac{\hbar^2\nabla^2}{2m_\uparrow}-\mu_\uparrow&\Delta_{\text{LO}}(z)\\
\Delta_{\text{LO}}(z)&\frac{\hbar^2\nabla^2}{2m_\downarrow}+\mu_\downarrow\end{array}\right)\left(\begin{array}{c}u_{n}({\bf x})\\v_{n}({\bf x})\end{array}\right)=\hbar\omega_{n}\left(\begin{array}{c}u_{n}({\bf x})\\v_{n}({\bf x})\end{array}\right),
\label{Bog}
\end{align}
where $u$ and $v$ are the quasiparticle coherence factors and $n$ is the band index labeling the energy eigenmode. The inverse Green's function, on the left-hand side of the equation, contains the Larkin-Ovchinnikov order parameter. Since this pairing field has a periodicity the quasiparticle wavefunctions are according to Bloch's theorem of the form
\begin{align}
\left(\begin{array}{c}u_{n}({\bf x})\\v_{n}({\bf x})\end{array}\right)=\left(\begin{array}{c}u_{{\bf k},n}(z)\\v_{{\bf k},n}(z)\end{array}\right)e^{\text{i}{\bf k\cdot x}},
\end{align}
where $u_{{\bf k},n}(z)$ and $v_{{\bf k},n}(z)$ are periodic functions with the same periodicity as the gap $\Delta_{\text{LO}}(z)$ and are thus only periodic in the $z$-direction, which is reflected also in the single band index $n$.

In order to determine the transition line between the LO and LO$^3$ phase we first obtain the equilibrium state of the thermodynamic potential for the LO phase and this we use as input to calculate $\alpha_\perp$. The line where $\alpha_\perp$ is zero is the transition line between the two inhomogeneous superfluid phases, see Figs. \ref{fasediagram}(a) and \ref{fasediagram2}.

\section{Inhomogeneous Densities}
\label{section densities}
\begin{figure}
\includegraphics[width=1.0\columnwidth]{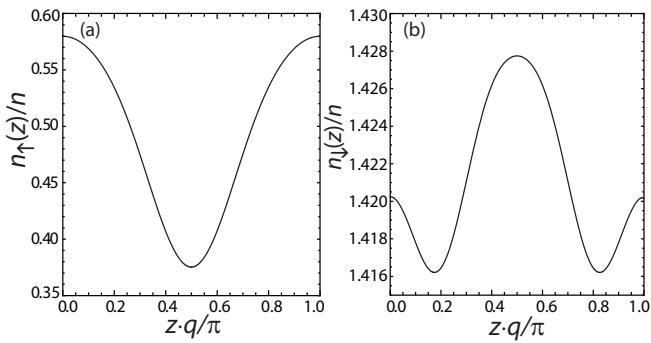}
\caption{\label{densities}
Densities for the ${}^6$Li atoms (a) and the ${}^{40}$K atoms (b) scaled by the total average density $n=[n_\uparrow(0)+n_\downarrow(0)]/2$ in the LO phase as a function of position for the point denoted with a cross in Fig. \ref{fasediagram}(a).
}
\end{figure}
From solving the Bogoliubov-de Gennes equation in Eq. (\ref{Bog}) we know the quasiparticle coherence factors $u$ and $v$ for the LO phase.
With these wavefunctions it is possible to calculate the particle densities for the ${}^6$Li and the ${}^{40}$K atoms as a function of position. Namely, the densities are given by
\begin{align}
\nonumber n_\sigma(z)=\sum_{{\bf k},n}&\left\{u^2_{{\bf k},n}(z)N_{\sigma}(\hbar\omega_{{\bf k},n})\right.\\
&\left.+v^2_{{\bf k},n}(z)[1-N_{-\sigma}(\hbar\omega_{{\bf k},n})]\right\}.
\end{align}
We take the vector ${\bf k}$ to lie in the first Brioullin zone in momentum space, which is here infinitely large in the $x$ and $y$ direction and has size $2q$ in the $z$ direction.
In Fig. \ref{densities} the inhomogeneous densities in the LO phase are shown for the point denoted with a cross in the phase diagram of Fig. \ref{fasediagram}(a) and it can be seen that the density of the minority atoms follows the form of the Cooper-pair ansatz. In
contrast, the density of the $^{40}$K atoms has its maximum
at the position where the Cooper pair has a minimum. In
this way the system relaxes the frustration caused by the
population imbalance, instead of phase separating, which
occurs for a majority of $^6$Li atoms [15, 16]. We find that
the modulation in the density of the $^6$Li atoms is about
10\%, as can be seen in Fig. \ref{densities}(a), whereas the modulation for the majority atoms is much smaller. The density modulations can be visible in an experiment, for example using Bragg spectroscopy. A homogeneous superfluid
does not show Bragg peaks and thus the appearance of
any Bragg peak is a sign of inhomogeneities in the superfluid phase.

\section{Discussion and Conlusion}
\label{section discussion and conclusion}
For the mixture of resonantly interacting ${}^6$Li and  ${}^{40}$K atoms we completed the phase diagram, which contains inhomogeneous superfluid phases. These superfluid phases have a crystalline order, just as a solid, and are thus special kinds of supersolids. Usually a supersolid is defined as a phase with both long-range diagonal and off-diagonal order in the one-particle density matrix. The LO and LO$^3$ only have diagonal order in the one-particle density matrix but there is both long-range diagonal and off-diagonal order in the two-particle density matrix. This means that the inhomogeneous superfluid is a supersolid phase, but not in the usual sense.

The results presented in this paper were calculated making use of two approximations.
The first one is the mean-field approximation, which is known to give good qualitative results. Although our results are only reasonable estimates quantitatively, we are confident that, as in the mass-balanced case, the phase diagram does not change qualitatively if screening and selfenergy effects are included.
However, fluctuations can have a large effect on the density modulations. In order to observe a spatial periodicity in the densities true long-range order is most desirable. But due to fluctuation effects only algebraic long-range order will exist in the LO phase \cite{Leo,Radzihovsky}, making it more difficult to observe the periodicity.

The second approximation we make is that we take into account only a single magnitude of the lattice wavevector $q$. For the continuous transition from the normal gas to the superfluid state this is exact, whereas it is an approximation inside the superfluid phase. Deep in the superfluid phase the pairing field is expected to take a more complicated form and to depend on a range of wave vectors \cite{Bulgac}.
If in the phase diagram the chemical-potential difference is on the $x$-axis all transitions take place close to the normal to superfluid transition as shown in Fig. \ref{fasediagram2}. Therefore, it is not expected that taking into account a more complicated form for the gap will change the phase diagram qualitatively. Moreover, our calculation is variational and thus gives a conservative estimate of the supersolid region of the phase diagram.

In this paper we presented results for a homogeneous system and we did not consider a confining trap. To take into account the trap for the atoms, a local-density approximation can be made.
The trap felt by the atoms depends on the species of the atoms, which means that for two atom species two different traps have to be taken into account \cite{Grimm1}.
The trap can have an advantageous influence. Namely, if the periodicity of the Cooper pair wavefunction is pinned by the shape of the trap it becomes easier to observe density modulations.

This work is supported by the Stichting voor Fundamenteel Onderzoek der Materie
(FOM), the Nederlandse Organisatie voor Wetenschaplijk Onderzoek (NWO).

\end{document}